\documentclass[12pt]{article}
\usepackage[T1]{fontenc}
\usepackage{amsfonts,amsmath}
\usepackage{epsfig,graphics}

\renewcommand{\appendix}{%
\renewcommand{\section}{%
\newpage
\thispagestyle{plain}%
\secdef\Appendix\sAppendix}%
\setcounter{section}{0}%
\renewcommand{\thesection}{\Alph{section}}%
}
\newcommand{\Appendix}[2][?]{%
\refstepcounter{section}%
\addcontentsline{toc}{Addendum}%
{\protect\numberline{\appendixname~\thesection}#1}%
{\flushleft\LARGE\bfseries\appendixname\ \thesection\par
\centering#2\par}%
\sectionmark{#1}\vspace{\baselineskip}}
\newcommand{\sAppendix}[1]{%
{\flushright\large\bfseries\appendixname\par
\centering#1\par}%
\vspace{\baselineskip}}

\def\be{\begin{equation}}

\def\bea{\begin{eqnarray}}

\def\eea{\end{eqnarray}}

\normalbaselines \oddsidemargin 0.5cm \evensidemargin 0.5cm
\begin{document}
%\doublespacing
\pagestyle{empty}
\begin{center}
{\Huge {\textbf{Entangled states, Lorentz transformations,
Spin-precession in magnetic fields}}}

\vspace{4mm} {\bf \large A. Chakrabarti\footnote{Email:
chakra@cpht.polytechnique.fr}}\\
 \vspace{2mm} \emph{Centre de
Physique Th{\'e}orique, Ecole Polytechnique, 91128 Palaiseau
Cedex, France.}

\end{center}
\begin{abstract}
{\small \noindent Two positive mass, spin $\frac 12$ particles
created in an entangled state are studied in the presence of a
constant magnetic field inducing distinct precessions, depending
on the respective momenta, of the two spins. The charge and
anomalous magnetic moment of each particle is taken into account.
Consequences for entanglement and, more generally, on
correlations, are derived. We start, however, with a compact
derivation of the effects of Lorentz transformations on such
entangled states, though that has been studied by several authors.
Our formalism displays conveniently the analogies and the
differences between the two cases. Moreover, combining the two,
one obtains the case of constant, orthogonal electric and magnetic
fields. More general perspectives are evoked in the concluding
remarks.}
\end{abstract}

\pagestyle{plain} \setcounter{page}{1}

%\tableofcontents

%\newpage

\section{Introduction}
\setcounter{equation}{0}

In studying entangled quantum states the particles involved are
usually assumed to move freely outward from their point of
production up to the apparatus, say Stern-Gerlach, where their
polarizations are measured. We refer to the review of GHSZ
\cite{1} where a full, lucid discussion can be found with ample
references to original sources. Two recent sources exploring
implications of non-locality and reality provide further insight
and, again, ample references \cite{2,3}.

Here we explore consequences on the correlations observed of
accelerations (Lorentz transformations) and of external constant
magnetic fields. Combining these two we obtain also effects of a
"crossed" electric field, meaning fields
$\left(\overrightarrow{\textbf{E}},\;\overrightarrow{\textbf{B}}\right)$,
constant and orthogonal, satisfying
\begin{equation}
\overrightarrow{\textbf{E}}\cdot\overrightarrow{\textbf{B}}=0,\qquad
\left|\overrightarrow{\textbf{E}}\right|<\left|\overrightarrow{\textbf{B}}\right|.
\end{equation}
We consider only massive $\left(m>0\right)$ spin-$\frac 12$
particle pairs with spin projections $\left(\pm \frac 12\right)$.
There are several recent studies of entanglement in the context of
Lorentz transformations \cite{4,5,6,7,8}. We summarize however
certain features (Sec.3) implementing our parametrization of
Wigner rotations (App. A). This turns out to be particularly
helpful for our study of entangled states in a magnetic field.
(Sec.4).

How does the spin react to the Lorentz transformations and
magnetic fields? The answer is -through "Wigner rotations". A
review \cite{9} presents systematically our relevant results,
referring to our previous papers (which will not be cited here
separately). A comparative study, citing a very large number of
sources, is provided by a recent review \cite{10}. Wigner's
construction of unitary representations of the Poincar\'e group
leads to irreducible ones labeled by two invariants, one with
continuous spectrum and the other with a discrete, integer or
half-integer one. The first one is the mass $\left(m\right)$ and
the second one spin $\left(s\right)$. (The zero mass, continuous
spin case does not seem to be realized in nature.) This is a
profound result. The purely group theoretic construction leads
directly to these two fundamental physical properties of a
particle.

How do spins, thus obtained, behave? Wigner's construction again
furnishes the answer in a canonical fashion for all spins. One
thus arrives at Wigner rotations for spins (App. A). Under pure
rotations the spin and the momentum turn about the same axis
through the same angle. Under pure Lorentz transformations, for
$m>0$, the spin turns again about the same axis but through a
smaller angle than the momentum. There is a a mass-dependent
"lag-effect". (For $m=0$, the spin catches up. The helicity
remains constant.) The precession of polarization in a magnetic
field, the full Thomas equation, can be obtained by starting with
the time derivative of Wigner rotations \cite{9}. In the following
sections the consequences of Lorentz transformations and
precessions for entangled states are studied systematically,
implementing such an approach.

\section{Recapitulation of basic features}
\setcounter{equation}{0}

Here, before even coming to Lorentz transformations, we summarize some well-known fundamental facts concerning two
massive $\left(m>0\right)$, spin-$\frac 12$ particles in an
entangled state. Rather than citing the famous original sources we
refer conveniently to the Apps. A, B, C,... of ref. 1. Let
\begin{equation}
\left|\psi\right\rangle=\frac
1{\sqrt{2}}\left(\left|+\right\rangle_1\left|-\right\rangle_2-\left|-\right\rangle_1\left|+\right\rangle_2\right)
\end{equation}
denote the state of total spin zero of two spin $\frac 12$
particles, of 4-velocities
$\left(u_0,\overrightarrow{\mathbf{u}}\right)$,
$\left(u_0,-\overrightarrow{\mathbf{u}}\right)$ respectively in
the frame of a particular observer (considered as the rest frame).
The following facts are well-known:
\begin{enumerate}
    \item  The state is "entangled". Namely, the assumption that one can express it as
\begin{equation}
\left|\psi\right\rangle=\left|\omega\right\rangle_1\left|z\right\rangle_2,
\end{equation}
where $\left|\omega\right\rangle_1$ and $\left|z\right\rangle_2$
are each a linear superposition of states in space of states of
particles $1$ and $2$ respectively, leads to a contradiction.

\item Rotating the axis of spin projections simultaneously
for the two particles to any direction $\widehat{\mathbf{n}}$
(with $\widehat{\mathbf{n}}^2=1$) leaves the righthand side of
(2.1) invariant. This is an expression of a spherical symmetry of
the spin zero state and holds as long as the axis of projection is
the same
$\left(\widehat{\mathbf{n}}_1=\widehat{\mathbf{n}}_2=\widehat{\mathbf{n}}\right)$
for the two particles.

\item When the two axes are different $\left(\widehat{\mathbf{n}}_1\neq
\widehat{\mathbf{n}}_2\right)$, taking (App. B of ref. 1)
$\widehat{\mathbf{n}}_2$ as the polar axis and
$\widehat{\mathbf{n}}_1$ with polar and azimuthal angles
$\left(\theta,0\right)$ respectively one obtains
\begin{eqnarray}
&&\left|\psi\right\rangle=\frac
1{\sqrt{2}}\left(-\sin\left(\frac\theta 2\right)
\left|\widehat{\mathbf{n}}_1,+\right\rangle_1\left|\widehat{\mathbf{n}}_2,+\right\rangle_2+
\cos\left(\frac\theta 2\right)
\left|\widehat{\mathbf{n}}_1,+\right\rangle_1\left|\widehat{\mathbf{n}}_2,-\right\rangle_2-\right.\nonumber\\
&&\phantom{\left|\psi\right\rangle=} \left.\cos\left(\frac\theta
2\right)
\left|\widehat{\mathbf{n}}_1,-\right\rangle_1\left|\widehat{\mathbf{n}}_2,+\right\rangle_2-
\sin\left(\frac\theta 2\right)
\left|\widehat{\mathbf{n}}_1,-\right\rangle_1\left|\widehat{\mathbf{n}}_2,-\right\rangle_2
\right).
\end{eqnarray}
The amplitudes for the joint outcomes are respectively ($\pm$
denoting the projections detected)
\begin{equation}
\left(P_{++},P_{+-},P_{-+},P_{--}\right)=\frac 12
\left(\sin^2\left(\frac\theta 2\right), \cos^2\left(\frac\theta
2\right), \cos^2\left(\frac\theta 2\right),\sin^2\left(\frac\theta
2\right)\right).
\end{equation}
consistent with $P_{++}+P_{+-}+P_{-+}+P_{--}=1$. Hence the
expectation value of the product of the measurement outcomes is
now
\begin{equation}
E\left(\widehat{\mathbf{n}}_1,\widehat{\mathbf{n}}_2\right)=P_{++}+P_{--}-P_{-+}-P_{+-}=\sin^2\left(\frac\theta
2\right)-\cos^2\left(\frac\theta 2\right)=-\cos\theta
=-\left(\widehat{\mathbf{n}}_1\cdot\widehat{\mathbf{n}}_2\right).
\end{equation}
For
$\widehat{\mathbf{n}}_1=\widehat{\mathbf{n}}_2=\widehat{\mathbf{n}}$
one obtains the perfect correlation ($\theta=0$) as
\begin{equation}
E=-1.
\end{equation}
[This is also called "anti-correlation" since for $\theta=0$ one
obtain $\left(+,-\right)$ or $\left(-,+\right)$ projections only.]

\item For measurements involving three directions, $\left(\widehat{\mathbf{n}}_1, \widehat{\mathbf{n}}_2,
\widehat{\mathbf{n}}_3\right)$ say, Bell's inequality can be
expressed in our notation as
\begin{equation}
\left|E\left(\widehat{\mathbf{n}}_1,\widehat{\mathbf{n}}_2\right)-E\left(\widehat{\mathbf{n}}_1,\widehat{\mathbf{n}}_3\right)\right|-
E\left({\mathbf{n}}_2,{\mathbf{n}}_3\right)\leq 1.\end{equation}
Bell derived this (in 1964) in the context of the argument of
Einstein, Podolski and Rosen (EPR) involving locality, reality and
completeness. (See, for example Sec. 2 of Ref. 1 and corresponding
references.) It is a famous fact that quantum mechanical
predictions can violate Bell's inequality. In particular,
implementing (2.5), representing such predictions, the l.h.s of
(2.7) becomes
\begin{equation}
\left|-\left(\widehat{\mathbf{n}}_1\cdot\widehat{\mathbf{n}}_2\right)+\left(\widehat{\mathbf{n}}_1\cdot\widehat{\mathbf{n}}_3\right)\right|+
\left(\widehat{\mathbf{n}}_2\cdot\widehat{\mathbf{n}}_3\right).
\end{equation}
Choosing, for example,
$\left(\widehat{\mathbf{n}}_1,\widehat{\mathbf{n}}_2,\widehat{\mathbf{n}}_3\right)$
in the $xy$-plane with azimuthal angles $\left(0,\frac\pi
3,\frac{2\pi}3\right)$ respectively one obtains
\begin{equation}
\left|-\left(\widehat{\mathbf{n}}_1\cdot\widehat{\mathbf{n}}_2\right)+\left(\widehat{\mathbf{n}}_1\cdot\widehat{\mathbf{n}}_3\right)\right|+
\left(\widehat{\mathbf{n}}_2\cdot\widehat{\mathbf{n}}_3\right)=\frac
32>1.
\end{equation}
\end{enumerate}

\section{Lorentz transformations and correlations}
\setcounter{equation}{0}
 Now we come to Lorentz transformations. Here, as in Sec.4, we restrict our considerations to eigenstates of equal
and opposite initial momenta for the two particles. Density
matrices are not introduced, as compared to some relevant recent
sources already cited. The transitions between the results of
Secs.2,3,4 respectively are thus displayed conveniently and
clearly. Consider an observer for whom the first frame (sec. 2) is
related through a pure Lorentz transformation corresponding to a
4-velocity $u''$. We now implement the result of App. A. To start
with the 4-velocities of particle 1 and particle 2 (of (2.2)) are
\begin{equation}
\left(u_0,\overrightarrow{\mathbf{u}}\right)\qquad
\hbox{and}\qquad \left(u_0,-\overrightarrow{\mathbf{u}}\right)
\end{equation}
respectively. Using (A.1), (A.2) define the corresponding
\begin{equation}
a_1=\left(1+u_0\right)
\left(1+u''_0\right)\left(1+u_0u''_0+\overrightarrow{\mathbf{u}}\cdot
\overrightarrow{\mathbf{u}}''\right),\qquad b_1=
1+u_0+u''_0+u_0u''_0+\overrightarrow{\mathbf{u}}\cdot
\overrightarrow{\mathbf{u}}''
\end{equation}
and
\begin{equation}
a_2=\left(1+u_0\right)
\left(1+u''_0\right)\left(1+u_0u''_0-\overrightarrow{\mathbf{u}}\cdot
\overrightarrow{\mathbf{u}}''\right),\qquad b_2=
1+u_0+u''_0+u_0u''_0-\overrightarrow{\mathbf{u}}\cdot
\overrightarrow{\mathbf{u}}''.
\end{equation}
The Wigner rotations of the spins (for
$\overrightarrow{\mathbf{u}}\times \overrightarrow{\mathbf{u}}''
\neq 0$) will be respectively about the axes
\begin{equation}
\widehat{\mathbf{k}}_{(1)}=-\widehat{\mathbf{k}}_{(2)}=\widehat{\mathbf{k}}=\frac{\overrightarrow{\mathbf{u}}\times
\overrightarrow{\mathbf{u}}''}{\left|\overrightarrow{\mathbf{u}}\times
\overrightarrow{\mathbf{u}}''\right|}
\end{equation}
through angles $\left(\delta_1,\delta_2\right)$ where
\begin{equation}
\cos\left(\frac{\delta_1} 2\right)=\frac{b_1}{\sqrt{2a_1}},\qquad
\cos\left(\frac{\delta_2} 2\right)=\frac{b_2}{\sqrt{2a_2}}.
\end{equation}
The spin states transform as
\begin{eqnarray}
&&\left(\begin{matrix}
  \left|+\right\rangle \\
  \left|-\right\rangle \\
\end{matrix}\right)_{(1)}\longrightarrow
e^{\mathbf{i}\delta_1\widehat{k}\cdot
\frac{\overrightarrow{\sigma}}{2}}\left(\begin{matrix}
  \left|+\right\rangle \\
  \left|-\right\rangle \\
\end{matrix}\right)_{(1)}\\
&&\left(\begin{matrix}
  \left|+\right\rangle \\
  \left|-\right\rangle \\
\end{matrix}\right)_{(2)}\longrightarrow
e^{-\mathbf{i}\delta_2\widehat{k}\cdot
\frac{\overrightarrow{\sigma}}{2}}\left(\begin{matrix}
  \left|+\right\rangle \\
  \left|-\right\rangle \\
\end{matrix}\right)_{(2)}.
\end{eqnarray}
(Here $\overrightarrow{\sigma}$ denote the Pauli matrices.) With
direction cosines of $\widehat{\textbf{k}}$ denoted as
$\left(k_1,k_2,k_3\right)$ satisfying
\begin{equation}
k_1^2+k_2^2+k_3^2=\widehat{\textbf{k}}^2=1
\end{equation}
one obtains for the entangled state (2.1)
\begin{eqnarray}
&&\frac
1{\sqrt{2}}\left(\left|+\right\rangle_1\left|-\right\rangle_2-\left|-\right\rangle_1\left|+\right\rangle_2\right)\longrightarrow
\frac1{\sqrt{2}}\left(\cos\frac
{\left(\delta_1+\delta_2\right)}2\left(\left|+\right\rangle_1\left|-\right\rangle_2-\left|-\right\rangle_1\left|+\right\rangle_2\right)+\right.\\
&&\left.\mathbf{i}\sin\frac
{\left(\delta_1+\delta_2\right)}2\left[k_3\left(\left|+\right\rangle_1\left|-\right\rangle_2+\left|-\right\rangle_1\left|+\right\rangle_2\right)
-\left(k_1+\mathbf{i}k_2\right)\left|+\right\rangle_1\left|+\right\rangle_2+\left(k_1-\mathbf{i}k_2\right)
\left|-\right\rangle_1\left|-\right\rangle_2\right]
\right)\nonumber
\end{eqnarray}
(For a rotation,one just sets $\delta_2=-\delta_1$. A rotation
$-\delta_1$ about $-\widehat{\textbf{k}}$ is one of $\delta_1$
about $\widehat{\textbf{k}}$. One thus recovers the rotational
invariance of the l.h.s. as mentioned in sec. 2, subsection (2).)

Fundamental consequences: We state them directly to start with
\begin{description}
\item[(1)] \underline{Entanglement is frame-independent}: Lorentz
transformations do not "disentangle" an entangled state. It
remains entangled in all frames.
\item[(2)] \underline{Violation of Bell's
inequality is frame-dependent}: For suitable choices of parameters
a violation of the inequality for the first observer can be absent
for the second after a Lorentz transformation.
\end{description}

We now demonstrate statement (1) directly. Then we extract
consequences of a Lorentz transformation concerning correlation.
This will lead to (2) along with other results.

Assume that the r.h.s. of (3.9) can be expressed (as in (2.2)) in
a non-entangled form as
\begin{equation}
\left(\mathrm{x}_1\left|+\right\rangle_1+
\mathrm{y}_1\left|-\right\rangle_1\right)
\left(\mathrm{x}_2\left|+\right\rangle_2+
\mathrm{y}_2\left|-\right\rangle_2\right).
\end{equation}
Comparing with (3.9) one obtains
\begin{eqnarray}
&&\mathrm{x}_1\mathrm{x}_2=-\mathbf{i}\left(k_1+\mathbf{i}k_2\right)\sin\frac
{\left(\delta_1+\delta_2\right)}2,\nonumber\\
&&\mathrm{y}_1\mathrm{y}_2=\mathbf{i}\left(k_1-\mathbf{i}k_2\right)\sin\frac
{\left(\delta_1+\delta_2\right)}2,\nonumber\\
&&\mathrm{x}_1\mathrm{y}_2=\cos\frac
{\left(\delta_1+\delta_2\right)}2+\mathbf{i}k_3\sin\frac
{\left(\delta_1+\delta_2\right)}2,\nonumber\\
&&\mathrm{x}_2\mathrm{y}_1=-\cos\frac
{\left(\delta_1+\delta_2\right)}2+\mathbf{i}k_3\sin\frac
{\left(\delta_1+\delta_2\right)}2.
\end{eqnarray}
This implies
\begin{equation}
\mathrm{x}_1\mathrm{x}_2\mathrm{y}_1\mathrm{y}_2=\left(k_1^2+k_2^2\right)\sin^2\frac
{\left(\delta_1+\delta_2\right)}2=-\cos^2\frac
{\left(\delta_1+\delta_2\right)}2-k_3^2\sin^2\frac
{\left(\delta_1+\delta_2\right)}2
\end{equation}
or, since (3.8) holds,
\begin{equation}
\cos^2\frac {\left(\delta_1+\delta_2\right)}2+\sin^2\frac
{\left(\delta_1+\delta_2\right)}2=0.
\end{equation}
But, from our construction, $\left(\delta_1,\delta_2\right)$ are
real angles (see (3.5)) and hence
\begin{equation}
\cos^2\frac {\left(\delta_1+\delta_2\right)}2+\sin^2\frac
{\left(\delta_1+\delta_2\right)}2=1.
\end{equation}
Thus one arrives at a contradiction. Hence statement (1) holds.

Let us now consider correlations as observed in the frame of the
second observer. To simplify notations set, in (3.9),
\begin{equation}
\cos\frac {\left(\delta_1+\delta_2\right)}2\equiv
\mathrm{c},\qquad \sin\frac
{\left(\delta_1+\delta_2\right)}2\equiv \mathrm{s}
\end{equation}
and introduce
\begin{equation}
\cos\frac{\theta'}2\equiv \mathrm{c}',\qquad
\sin\frac{\theta'}2\equiv \mathrm{s}'
\end{equation}
for the new polar angle of the spin projections of particle 2 with
the corresponding states
\begin{equation}
\left|+\right\rangle_2=\mathrm{c}'\left|+\right\rangle'_2+\mathrm{s}'
\left|-\right\rangle_2',\qquad
\left|-\right\rangle_2=-\mathrm{s}'\left|+\right\rangle'_2+\mathrm{c}'
\left|-\right\rangle_2'.
\end{equation}
The r.h.s. of (3.9) is then
\begin{equation}
\frac
1{\sqrt{2}}\left(\alpha_1\left|+\right\rangle_1\left|+\right\rangle'_2+
\alpha_2\left|+\right\rangle_1\left|-\right\rangle'_2+
\beta_1\left|-\right\rangle_1\left|+\right\rangle'_2+
\beta_2\left|-\right\rangle_1\left|-\right\rangle'_2\right),
\end{equation}
where
\begin{eqnarray}
&&\alpha_1=\left(k_2\mathrm{sc}'-\mathrm{cs}'\right)-\mathbf{i}\mathrm{s}\left(k_1\mathrm{c}'+k_3\mathrm{s}'\right),\qquad
\beta_2=\left(k_2\mathrm{sc'-cs}'\right)+\mathbf{i}\mathrm{s}\left(k_1\mathrm{c}'+k_3\mathrm{s}'\right),\nonumber\\
&&\alpha_2=\left(k_2\mathrm{ss'+cc}'\right)-\mathbf{i}\mathrm{s}\left(k_1\mathrm{s}'-k_3\mathrm{c}'\right),\qquad
\beta_1=-\left(k_2\mathrm{ss'+cc}'\right)-\mathbf{i}\mathrm{s}\left(k_1\mathrm{s}'-k_3\mathrm{c}'\right)\;\phantom{xxx}.
\end{eqnarray}
The amplitude for the spin projections (compare (2.4)) are now
\begin{eqnarray}
&&P_{++}=P_{--}=\frac 12 \left(\left(k_2\mathrm{sc'-cs'}\right)^2+\mathrm{s}^2\left(k_1\mathrm{c}'+k_3\mathrm{s}'\right)^2\right),\nonumber\\
&&P_{+-}=P_{-+}=\frac 12
\left(\left(k_2\mathrm{ss'+cc}'\right)^2+\mathrm{s}^2\left(k_3\mathrm{c}'-k_1\mathrm{s}'\right)^2\right).
\end{eqnarray}
As a check one notes (using
$\mathrm{c^2+s^2}=\mathrm{{c'}^2+{s'}^2}=k_1^2+k_2^2+k_3^2=1$)
\begin{equation}
P_{++}+P_{--}+P_{+-}+P_{-+}=1.
\end{equation}
The correlation is now (compare (2.5))
\begin{eqnarray}
&&E=P_{++}+P_{--}-P_{+-}-P_{-+}\nonumber\\
&&\phantom{E}=
\left(k_2\mathrm{s(c'+s')+c(c'-s')}\right)\left(k_2\mathrm{s(c'-s')-c(c'+s')}\right)+\nonumber\\
&&\phantom{E=}
s^2\left(k_1\mathrm{(c'-s')}+k_3\mathrm{(c'+s')}\right)\left(k_1\mathrm{(c'+s')}-k_3\mathrm{(c'-s')}\right).
\end{eqnarray}
This varies with the parameters in a relatively complicated
fashion. Corresponding to particular choices of the axis of the
Wigner rotation $\left(\widehat{\textbf{k}}\approx
\overrightarrow{\mathbf{u}} \times
\overrightarrow{\mathbf{u}}"\right)$ one obtains simple particular
cases of interest.
\begin{description}
    \item[case 1.]
\begin{eqnarray}
&& k_1=k_2=0,\qquad k_3=\pm 1\\
&&E=-\left(\mathrm{{c'}^2-{s'}^2}\right)=-\cos\theta'.
\end{eqnarray}
    \item[Case 2.] \begin{eqnarray}
&& k_3=k_1=0,\qquad k_2=\pm 1\\
&&E=-\left(\left(\mathrm{c^2-s^2}\right)\left(\mathrm{{c'}^2-{s'}^2}\right)\pm
\left(\mathrm{2cs}\right)\left(\mathrm{2c's'}\right)\right)=-\cos\left(\delta_1+\delta_2\mp
\theta'\right).
\end{eqnarray}
For, respectively, $\delta_1+\delta_2=\pm \theta'$ the
(anti)correlation becomes perfect for the second observer.
\item[Case 3.]
\begin{eqnarray}
&& k_2=k_3=0,\qquad k_1=\pm 1\\
&&E=-\left(\mathrm{c^2-s^2}\right)\left(\mathrm{{c'}^2-{s'}^2}\right)=-\cos\left(\delta_1+\delta_2\right)\cos\theta'.
\end{eqnarray}
This last case suffices to illustrate the possibility presented as
statement (2) concerning the frame dependence of the inequality.
Varying $\theta'$ corresponding to $\left(\widehat{\textbf{n}}_1,
\widehat{\textbf{n}}_2, \widehat{\textbf{n}}_3\right)$ involved in
(2.7), (2.8) and (2.9) (namely our $\theta'$) the r.h.s. of (2.9)
becomes, (due to the extra factor in (3.25)) for the second
observer
\begin{equation}
\frac 32\cos\left(\delta_1+\delta_2\right).
\end{equation}
For
\begin{equation}
\cos\left(\delta_1+\delta_2\right)<\frac 23
\end{equation}
Bell's inequality is satisfied in the frame of the second observer
while it is violated in that of the first. We have thus
established its frame-dependence.
\end{description}

\section{Entangled state in a constant magnetic field}
\setcounter{equation}{0}

The equation for precession of canonical polarization operators
are presented in App. B. We recapitulate briefly the basic
equations and finals results. More details can be found in
appendix B. Let
$\left(\overrightarrow{\mathbf{B}},\overrightarrow{\mathbf{v}},\overrightarrow{\Sigma}\right)$
be respectively the constant, homogenous magnetic field, the
velocity and the polarization. With unit vectors
$\left(\widehat{\mathbf{B}},\widehat{\mathbf{v}}\right)$,
\begin{equation}
\overrightarrow{\textbf{B}}=B\widehat{\textbf{B}},\qquad
\overrightarrow{\textbf{v}}=v\widehat{\textbf{v}},\qquad
\gamma=\left(1-v^2\right)^{-1/2},\qquad \alpha=\left(g-2\right)/2
\end{equation}
and
\begin{eqnarray}
&&\overrightarrow{\omega}=\frac{eB}{m\gamma}\widehat{\textbf{B}},\qquad
\overrightarrow{\Omega}=\frac{\alpha
eB}{m\gamma}\left(\gamma\widehat{\textbf{B}}-\left(\gamma-1\right)\left(\widehat{\textbf{B}}\cdot
\widehat{\textbf{v}}\right)\widehat{\textbf{v}}\right)
\end{eqnarray}
the equations are
\begin{eqnarray}
&&\frac{d\overrightarrow{\mathbf{v}}}{dt}=-\overrightarrow{\omega}\times
\overrightarrow{\mathbf{v}},\qquad
\frac{d\overrightarrow{\Sigma}}{dt}=-\left(\overrightarrow{\omega}+\overrightarrow{\Omega}
\right)\times \overrightarrow{\Sigma}.
\end{eqnarray}
Apart from $\left(v,\widehat{\mathbf{B}}\cdot
\widehat{\mathbf{v}}\right)$ one has constants
\begin{eqnarray}
&&\omega^2=\left(\frac{eB}{m\gamma}\right)^2,\qquad
\Omega^2=\left(\frac{\alpha
eB}{m\gamma}\right)^2\left(\gamma^2-\left(\gamma^2-1\right)\left(\widehat{\textbf{B}}\cdot
\widehat{\textbf{v}}\right)^2\right)
\end{eqnarray}
where with unit vectors
$\left(\widehat{\omega},\widehat{\Omega}\right)$
$\overrightarrow{\omega}=\omega\widehat{\omega}$,
$\overrightarrow{\Omega}=\Omega\widehat{\Omega}$. For
\begin{equation}
0\leq\left(\widehat{\textbf{B}}\cdot
\widehat{\textbf{v}}\right)^2<1
\end{equation}
one starts with ((0) denoting initial value at $t=0$)
ortho-normalized fixed axes
\begin{equation}
\widehat{\textbf{B}}, \qquad
\frac{\left(\widehat{\textbf{B}}\times
\widehat{\textbf{v}}\right)_{(0)}}{\sqrt{1-\left(\widehat{\textbf{B}}\cdot
\widehat{\textbf{v}}\right)^2}},\qquad
\frac{\widehat{\textbf{B}}\times\left(\widehat{\textbf{B}}\times
\widehat{\textbf{v}}\right)_{(0)}}{\sqrt{1-\left(\widehat{\textbf{B}}\cdot
\widehat{\textbf{v}}\right)^2}},
\end{equation}
Then one defines ortho-normalized axes rotating about
$\widehat{\mathbf{B}}$ as (using values at time $t$)
\begin{equation}
\widehat{\textbf{B}}, \qquad
\frac{\left(\widehat{\textbf{B}}\times
\widehat{\textbf{v}}\right)}{\sqrt{1-\left(\widehat{\textbf{B}}\cdot
\widehat{\textbf{v}}\right)^2}},\qquad
\frac{\widehat{\textbf{B}}\times\left(\widehat{\textbf{B}}\times
\widehat{\textbf{v}}\right)}{\sqrt{1-\left(\widehat{\textbf{B}}\cdot
\widehat{\textbf{v}}\right)^2}},
\end{equation}
The projections of $\overrightarrow{\Omega}$ are
\begin{eqnarray}
&&\Omega_1=\widehat{\textbf{B}}\cdot \overrightarrow{\Omega}=
\frac{\alpha
eB}{m\gamma}\left(\gamma-\left(\gamma-1\right)\left(\widehat{\textbf{B}}\cdot
\widehat{\textbf{v}}\right)^2\right),\nonumber \\
&&\Omega_2=\frac{\left(\widehat{\textbf{B}}\times
\widehat{\textbf{v}}\right)\cdot\overrightarrow{\Omega}}{\sqrt{1-\left(\widehat{\textbf{B}}\cdot
\widehat{\textbf{v}}\right)^2}}=0,\nonumber\\
&&\Omega_3=
\frac{\left(\widehat{\textbf{B}}\times\left(\widehat{\textbf{B}}\times
\widehat{\textbf{v}}\right)\right)\cdot\overrightarrow{\Omega}}{\sqrt{1-\left(\widehat{\textbf{B}}\cdot
\widehat{\textbf{v}}\right)^2}}= \frac{\alpha
eB}{m\gamma}\left(\gamma-1\right)\left(\widehat{\textbf{B}}\cdot
\widehat{\textbf{v}}\right)\sqrt{1-\left(\widehat{\textbf{B}}\cdot
\widehat{\textbf{v}}\right)^2},
\end{eqnarray}
consistent with $\Omega_1^2+\Omega_3^2=\Omega^2$. The solutions
(presented in detail in App. B) combines successive rotations of
$\overrightarrow{\Sigma}$ through \begin{enumerate}
    \item an angle $\left(\omega t\right)$ about the axis
    $-\widehat{\omega}(=-\widehat{\mathbf{B}})$
    \item an angle $\left(\Omega t\right)$ about the axis
    $-\widehat{\Omega}$
\end{enumerate}
From (B.8) it follows that in passing from particle 1 with initial
velocity $\overrightarrow{\mathbf{v}}_{(0)}$ to particle 2 with
$-\overrightarrow{\mathbf{v}}_{(0)}$ is equivalent to
\begin{equation}
\omega t\longrightarrow \omega t+\pi.
\end{equation}
Thus the transformation matrices for the spin-$\frac 12$ states
$\left(\left|+\right\rangle, \left|-\right\rangle\right)$ are
respectively
\begin{equation}
e^{-\mathbf{i}\frac{\Omega t}2\left(\widehat{\Omega}\cdot
\overrightarrow{\sigma}\right)}e^{-\mathbf{i}\frac{\omega
t}2\left(\widehat{\mathbf{B}}\cdot \overrightarrow{\sigma}\right)}
\end{equation}
for particle 1, and
\begin{equation}
e^{-\mathbf{i}\frac{\Omega t}2\left(\widehat{\Omega}\cdot
\overrightarrow{\sigma}\right)}e^{-\mathbf{i}\frac{\left(\omega
t+\pi\right)}2\left(\widehat{\mathbf{B}}\cdot
\overrightarrow{\sigma}\right)}
\end{equation}
for particle 2. Here
\begin{equation}
\widehat{\Omega}\cdot \overrightarrow{\sigma}=\frac
1\Omega\left(\Omega_1\sigma_1+\Omega_3\sigma_3\right).
\end{equation}
For $\left(\widehat{\mathbf{B}}\cdot
\widehat{\mathbf{v}}\right)=0$ the two axes coincide and one has
simply the matrices
\begin{equation}
e^{-\mathbf{i}\left(\frac{\Omega+\omega}2\right)
t\left(\widehat{\mathbf{B}}\cdot \overrightarrow{\sigma}\right)}
\end{equation}
and
\begin{equation}
e^{-\mathbf{i}\left(\frac{\left(\Omega+\omega\right)t+\pi}2\right)\left(\widehat{\mathbf{B}}\cdot
\overrightarrow{\sigma}\right)}
\end{equation}
for the particles 1 and 2 respectively. (N.B: The implementation
of (4.9) has already taken care of the effect of the change of
sign of $\overrightarrow{\mathbf{v}}_0$ on the axis of projection corresponding to
$\Omega_3$  through that on $\left(\widehat{\mathbf{B}}\times \widehat{\mathbf{v}}\right)$.
This is the content of (B.8). After this one should not include an
additional inversion of sign of the factor
$\left(\widehat{\mathbf{B}}\cdot \widehat{\mathbf{v}}\right)$ of
$\Omega_3$. This would amount to a double negative on one part and
lead to inconsistent results - such as time-dependent precession
of a spin zero-state.)

\begin{description}
    \item[case 1.] $\left(\widehat{\mathbf{B}}\cdot \widehat{\mathbf{v}}=0\right)$ We start the study of correlations with the
    simple case corresponding to (4.13). We keep, however, general
    direction cosines of $\overrightarrow{\mathbf{B}}$, so that
\begin{equation}
\widehat{\mathbf{B}}\cdot
\overrightarrow{\sigma}=\left(b_1\sigma_1+b_2\sigma_2+b_3\sigma_3\right).
\end{equation}
where, by definition,
\begin{equation}
b_1^2+b_2^2+b_3^2=1.
\end{equation}
Set $\mathrm{c}=\cos\frac 12\left(\Omega+\omega\right)t$,
$\mathrm{s}=\sin\frac 12\left(\Omega+\omega\right)t$. From (4.13),
the time-evaluation of particles 1 and 2 are respectively
\begin{eqnarray}
&&\left(\begin{matrix}
  \left|+\right\rangle_1 \\
  \left|-\right\rangle_1 \\
\end{matrix}\right)_{(t)}=
\begin{bmatrix}
  \mathrm{c}-\mathbf{i}b_3\mathrm{s} & -\mathbf{i}\left(b_1+\mathbf{i}b_2\right)\mathrm{s} \\
  -\mathbf{i}\left(b_1+\mathbf{i}b_2\right)\mathrm{s} & \mathrm{c}+\mathbf{i}b_3\mathrm{s} \\
\end{bmatrix}\left(\begin{matrix}
  \left|+\right\rangle_1 \\
  \left|-\right\rangle_1 \\
\end{matrix}\right)_{(0)}\\
&&\left(\begin{matrix}
  \left|+\right\rangle_2 \\
  \left|-\right\rangle_2 \\
\end{matrix}\right)_{(t)}=
\begin{bmatrix}
  -\mathrm{s}-\mathbf{i}b_3\mathrm{c} & -\mathbf{i}\left(b_1-\mathbf{i}b_2\right)\mathrm{c} \\
  -\mathbf{i}\left(b_1+\mathbf{i}b_2\right)\mathrm{c} & -\mathrm{s}+\mathbf{i}b_3\mathrm{c} \\
\end{bmatrix}\left(\begin{matrix}
  \left|+\right\rangle_2 \\
  \left|-\right\rangle_2 \\
\end{matrix}\right)_{(0)}.
\end{eqnarray}
Implementing $c^2+s^2=1$, $b_1^2+b_2^2+b_3^2=1$, one obtains
\begin{eqnarray}
&&\left(\left|+\right\rangle_1\left|-\right\rangle_2-
\left|-\right\rangle_1\left|+\right\rangle_2\right)_{(t)}=
-\mathbf{i}\left(b_1+\mathbf{i}b_2\right)\left|+\right\rangle_1\left|+\right\rangle_2+
\mathbf{i}\left(b_1-\mathbf{i}b_2\right)\left|-\right\rangle_1\left|-\right\rangle_2+\nonumber\\
&&\phantom{\left(\left|+\right\rangle_1\left|-\right\rangle_2-
\left|-\right\rangle_1\left|+\right\rangle_2\right)_{(t)}=}
\mathbf{i}b_3 \left(\left|+\right\rangle_1\left|-\right\rangle_2+
\left|-\right\rangle_1\left|+\right\rangle_2\right).
\end{eqnarray}
All time-dependence disappears (through
$\mathrm{c}^2+\mathrm{s}^2=1$) in the spin-zero entangled
state.This is consistent with the fact that a zero-spin does not
undergo precession. The structure on the right hand of (4.19)
should be compared to that of (2.3). But it remains entangled. The
demonstration is parallel to (3.10-14) (for change of frame).
Assume that one can express (4.19) as
\begin{equation}
\left(\mathrm{x}_1\left|+\right\rangle_1+\mathrm{y}_1\left|-\right\rangle_1\right)
\left(\mathrm{x}_2\left|+\right\rangle_2+\mathrm{y}_2\left|-\right\rangle_2\right).
\end{equation}
Then one must have
\begin{equation}
\mathrm{x}_1\mathrm{x}_2=-\mathbf{i}\left(b_1+\mathbf{i}b_2\right),\qquad
\mathrm{y}_1\mathrm{y}_2=\mathbf{i}\left(b_1-\mathbf{i}b_2\right),\qquad
\mathrm{x}_1\mathrm{y}_2=\mathbf{i}b_3,\qquad
\mathrm{x}_2\mathrm{y}_1=\mathbf{i}b_3
\end{equation}
and hence
$\mathrm{x}_1\mathrm{x}_2\mathrm{y}_1\mathrm{y}_2=b_1^2+b_2^2=-b_3^2$
or
\begin{equation}
b_1^2+b_2^2+b_3^2=0.
\end{equation}
This contradicts (4.16). Hence (4.19) represents an entangled
state.

{\bf Probabilities and correlations:} We again proceed as in sec.
3 (see (3.16-25)). Rotate the states of particle 2 as
\begin{equation}
\left|+\right\rangle_2\longrightarrow
\mathrm{c}'\left|+\right\rangle_2+\mathrm{s}'\left|-\right\rangle_2,\qquad
\left|-\right\rangle_2\longrightarrow
-\mathrm{s}'\left|+\right\rangle_2+\mathrm{c}'\left|-\right\rangle_2,
\end{equation}
where $\mathrm{c}'=\cos\left(\theta'/2\right)$,
$\mathrm{s}'=\sin\left(\theta'/2\right)$. Including a
normalization factor $\frac 1{\sqrt{2}}$, the. r.h.s. of (4.19) is
now
\begin{eqnarray}
&&\frac
{\mathbf{i}}{\sqrt{2}}\left[\left(-\left(b_1+ib_2\right)c'-b_3s'\right)
\left|+\right\rangle_1\left|+\right\rangle_2+
\left(\left(b_1-ib_2\right)c'+b_3s'\right)
\left|-\right\rangle_1\left|-\right\rangle_2-\right.\nonumber\\
&&\left. \left(\left(b_1+ib_2\right)s'-b_3c'\right)
\left|+\right\rangle_1\left|-\right\rangle_2-\left(\left(b_1-ib_2\right)s'-b_3c'\right)
\left|-\right\rangle_1\left|+\right\rangle_2\right]
\end{eqnarray}
The corresponding probabilities are
\begin{eqnarray}
&&P_{++}=P_{--}=\frac
1{2}\left(\left(b_1^2+b_2^2\right){c'}^2+b_3^2{s'}^2+2b_1b_3c's'\right),\nonumber\\
&& P_{+-}=P_{-+}=\frac
1{2}\left(\left(b_1^2+b_2^2\right){s'}^2+b_3^2{c'}^2-2b_1b_3c's'\right)
\end{eqnarray}
satisfying
\begin{equation}
P_{++}+P_{--}+
P_{+-}+P_{-+}=\left(b_1^2+b_2^2+b_3^2\right)\left({s'}^2+{c'}^2\right)=1.
\end{equation}
The correlation is
\begin{eqnarray}
&&E=P_{++}+P_{--}-
P_{+-}-P_{-+}=\left(b_1^2+b_2^2-b_3^2\right)\left({c'}^2-{s'}^2\right)+\left(2b_1b_3\right)\left(2c's'\right)\nonumber
\\
&&\phantom{E}=\left(b_1^2+b_2^2-b_3^2\right)\cos\theta'+\left(2b_1b_3\right)\sin\theta'\nonumber
\\
&&\phantom{E}=\cos\theta'-2b_3\left(b_3\cos\theta'-b_1\sin\theta'\right)
\end{eqnarray}
As in Sec. 3 one can now consider consequences for different
choices of the parameters. (Compare with (3.22-25).) Thus, for
example, setting
\begin{eqnarray}
&& b_2=0,\qquad b_3=\cos\left(\varphi/2\right),\qquad
b_1=\sin\left(\varphi/2\right),\\
&& E=-\cos\left(\varphi+\theta'\right).
\end{eqnarray}
For $b_3=0$,
\begin{equation}
E=\cos\theta'.
\end{equation}
For $b_1=0$, $b_3=\cos\left(\psi/2\right)$,
$b_2=\sin\left(\psi/2\right)$
\begin{equation}
E=\left(1-2b_3^2\right)\cos\theta'=-\cos\psi. \cos\theta'
\end{equation}
Consequences analogous to those of Sec. 3 can be observed. Thus
(compare (3.25-27)), for the same strength of the homogenous
magnetic field $\left(B\right)$ rotating it (changing $\psi$, say,
in (4.31)) one can satisfy or violate Bell's inequality, (for the
same
$\left(\widehat{\mathbf{n}}_1,\widehat{\mathbf{n}}_2,\widehat{\mathbf{n}}_3\right)$
as in the passage from (2.9) to (3.26)).

    \item[Case 2.] $\left(0<\left(\widehat{\mathbf{B}}\cdot \overrightarrow{\mathbf{v}}\right)^2<1\right)$
    We now have the general spinor matrices (4.10) and (4.11) for
    particles 1 and 2 respectively. Having illustrated the
    consequences of rotating the magnetic field $\overrightarrow{\mathbf{B}}$
    (i.e. variations of $\left(b_1,b_2,b_3\right)$ in the
    preceding subsection (for $\overrightarrow{\mathbf{B}}\cdot
    \overrightarrow{\mathbf{v}}=0$)  let us simplify the formalism
    here (for non-zero $\overrightarrow{\mathbf{B}}\cdot \overrightarrow{\mathbf{v}}$)
    by choosing axes such that (conserving (4.12))
\begin{equation}
\overrightarrow{\mathbf{B}}\cdot\overrightarrow{\sigma}=B\sigma_1,\qquad
\overrightarrow{\Omega}\cdot
\overrightarrow{\sigma}=\Omega_1\sigma_1+\Omega_3\sigma_3.
\end{equation}
For the limit $\left(\widehat{\mathbf{B}}\cdot
\overrightarrow{\mathbf{v}}\right)=0$ this corresponds to $b_1=1$,
$b_2=b_3=0$ leading in (4.19) to
\begin{equation}
\left(\left|+\right\rangle_1\left|-\right\rangle_2-
\left|-\right\rangle_1\left|+\right\rangle_2\right)_{(t)}=-\mathbf{i}\left(\left|+\right\rangle_1\left|+\right\rangle_2-
\left|-\right\rangle_1\left|-\right\rangle_2\right)_{(0)}
\end{equation}
and the contradiction between (4.16) and (4.22) reduces to that
between
\begin{equation}
b_1^2=1\qquad b_1^2=0.\end{equation} With notations
\begin{equation}
\frac{\Omega_1}{\Omega}=l_1,\qquad
\frac{\Omega_3}{\Omega}=l_3,\qquad l_1^2+l_3^2=1\end{equation} the
spinor matrices for particles 1 and 2 respectively are now
\begin{equation}
M_1=e^{-\mathbf{i}\frac{\Omega
t}2\left(l_1\sigma_1+l_3\sigma_3\right)}e^{-\mathbf{i}\frac{\omega
t}2\sigma_1},\qquad M_2=e^{-\mathbf{i}\frac{\Omega
t}2\left(l_1\sigma_1+l_3\sigma_3\right)}e^{-\mathbf{i}\frac{\left(\omega
t+\pi\right)}2\sigma_1}.
\end{equation}
Set
\begin{eqnarray}
&&\cos\left(\frac 12\omega t\right)=\mathrm{c},\qquad
\sin\left(\frac 12\omega t\right)=\mathrm{s},\nonumber\\
&&\cos\left(\frac 12\Omega t\right)=\mathrm{c}',\qquad
\sin\left(\frac 12\Omega t\right)=\mathrm{s}'
\end{eqnarray}
and
\begin{equation}
\alpha=\left(\mathrm{c'c}-l_1\mathrm{s's}\right)-\mathbf{i}l_3\mathrm{s'c},
\qquad
\beta=\left(\mathrm{c's}+l_1\mathrm{s'c}\right)-\mathbf{i}l_3\mathrm{s's}.
\end{equation}
Then
\begin{equation}
M_1=\begin{bmatrix}
  \alpha & -\mathbf{i}\beta \\
  -\mathbf{i}\beta^\ast & \alpha^\ast \\
\end{bmatrix}, \qquad
M_2=\begin{bmatrix}
  -\beta & -\mathbf{i}\alpha \\
  -\mathbf{i}\alpha^\ast & -\beta^\ast \\
\end{bmatrix}
\end{equation}
Thus through $\left(\alpha,\beta\right)$ are complicated,
$\left(M_1,M_2\right)$ have simple structures in their terms. From
(4.38) along with
$\mathrm{c}^2+\mathrm{s}^2=1=\mathrm{c'}^2+\mathrm{s'}^2$) one
obtains
\begin{equation}
\alpha\alpha^\ast+\beta\beta^\ast=\mathrm{c'}^2+\left(l_1^2+l_3^2\right)\mathrm{s'}^2=\mathrm{c'}^2+\mathrm{s'}^2=1.
\end{equation}
This implies the unitarity constraints
\begin{equation}
M_1^+M_1=M_2^+M_2=\begin{bmatrix}
  1 & 0 \\
  0 & 1 \\
\end{bmatrix}.
\end{equation}
Now
\begin{eqnarray}
&&\left(\left|+\right\rangle_1\left|-\right\rangle_2-
\left|-\right\rangle_1\left|+\right\rangle_2\right)_{(t)}=
\left[-\left(\alpha\left|+\right\rangle_1-i\beta\left|-\right\rangle_1\right)
\left(i\alpha^\ast\left|+\right\rangle_2+\beta^\ast\left|-\right\rangle_2\right)+\right.\nonumber\\
&&\phantom{\left(\left|+\right\rangle_1\left|-\right\rangle_2-
\left|-\right\rangle_1\left|+\right\rangle_2\right)_{(t)}=}\left.
\left(-i\beta^\ast\left|+\right\rangle_1+\alpha^\ast\left|-\right\rangle_1\right)
\left(\beta\left|+\right\rangle_2+i\alpha\left|-\right\rangle_2\right)\right]_{(0)}\nonumber\\
&&\phantom{\left(\left|+\right\rangle_1\left|-\right\rangle_2-
\left|-\right\rangle_1\left|+\right\rangle_2\right)_{(t)}}=
-\mathbf{i}\left(\alpha\alpha^\ast+\beta\beta^\ast\right)\left(\left|+\right\rangle_1\left|+\right\rangle_2-
\left|-\right\rangle_1\left|-\right\rangle_2\right)_{(0)}-\nonumber\\
&&\phantom{\left(\left|+\right\rangle_1\left|-\right\rangle_2-
\left|-\right\rangle_1\left|+\right\rangle_2\right)_{(t)}=}
\left(\alpha\beta^\ast+\alpha\beta^\ast\right)\left(\left|+\right\rangle_1\left|-\right\rangle_2\right)_{(0)}+
\nonumber\\
&&\phantom{\left(\left|+\right\rangle_1\left|-\right\rangle_2-
\left|-\right\rangle_1\left|+\right\rangle_2\right)_{(t)}=}
\left(-\beta\alpha^\ast+\alpha^\ast\beta\right)\left(\left|-\right\rangle_1\left|+\right\rangle_2\right)_{(0)}
\nonumber\\
&&\phantom{\left(\left|+\right\rangle_1\left|-\right\rangle_2-
\left|-\right\rangle_1\left|+\right\rangle_2\right)_{(t)}}
=-\mathbf{i}\left(\left|+\right\rangle_1\left|+\right\rangle_2-
\left|-\right\rangle_1\left|-\right\rangle_2\right)_{(0)}
\end{eqnarray}
Thus not only again time-dependence has disappeared for the
spin-zero entangled state but again for non-zero
$\left(\widehat{\mathbf{B}}\cdot\overrightarrow{\mathbf{v}}\right)$
the relation (4.33) (for $\left(\widehat{\mathbf{B}}\cdot
\overrightarrow{\mathbf{v}}=0\right)$, $b_2=b_3=0$) is reproduced.
Now one can rotate the axes to give more general orientations to
$\left(\widehat{\mathbf{B}},\widehat{\Omega}\right)$. The
consequences can be deduced in a straightforward fashion. We will
not go through the steps here.

\item[Case 3.] $\left(\widehat{\mathbf{B}}\cdot \overrightarrow{\mathbf{v}}=\pm 1\right)$
As discussed in App. B (B.25-26) this case has to be treated
separately. The normalization (4.6) is no longer well-defined. But
now $v$ is constant. Thus this case can, be treated quite simply
as a separate one. We will not present such a discussion here.
\end{description}

\section{Constant, orthogonal electric and magnetic fields}
\setcounter{equation}{0}

We only briefly indicate certain possibilities. Two cases have
been studied in our previous papers and solutions have been
obtained for the polarization. They are briefly presented in our
review (Ref. 9) with references to our original works.
\begin{description}
\item[Case 1: $\left(\left|\overrightarrow{\textbf{E}}\right|<\left|\overrightarrow{\textbf{B}}\right|\right)$]
This is presented in Sec. 3 of Ref. 9 (3.17-23). A Lorentz
transformation corresponding to the 4-velocity
\begin{equation}
u''=\frac 1{\sqrt{1-\frac{E^2}{B^2}}}\left(1,\frac
EB\;\widehat{\textbf{E}} \times \widehat{\textbf{B}}\right),
\end{equation}
where $\left(\widehat{\textbf{E}},\widehat{\textbf{B}}\right)$ are
unit vectors gives transformed fields
\begin{equation}
\overrightarrow{\textbf{E}}'=0,\qquad
\overrightarrow{\textbf{B}}'=
\left(1-\frac{E^2}{B^2}\right)^{-1/2}\overrightarrow{\textbf{B}}.
\end{equation}
Thus combining the results presented here in secs. 3 and 4 one can
analyze this case fully.
    \item[Case 2:
    $\left(\left|\overrightarrow{\textbf{E}}\right|=\left|\overrightarrow{\textbf{B}}\right|\right)$]
    (See (4.13-14) of Ref. 9 and sources cited there.) The
    Dirac  equation (with the Pauli term for anomalous magnetic
    moment and also electric dipole moment) has been solved for an
    external plane wave field. A particular limiting case
    corresponds to the present one. Complete solutions for
    polarization were obtained. The general plane wave will be
    studied elsewhere in the context of entanglement.
\end{description}

\section{Remarks}
\setcounter{equation}{0}

We have studied the effects of accelerations (Lorentz
transformations) and constant, homogenous magnetic fields on the
entangled state of two positive mass, spin $\frac 12$ particles.
An additional constant electric field orthogonal to the magnetic
one has also been briefly considered. We approach both aspects
systematically via Wigner-rotations of canonical spin,
recapitulated in Apps. A and B. Relevant basic features of
entanglement are, (correlations and Bell's Inequality)
recapitulated in sec. 2 (We consider both these domains as
well-known and refer to thorough review articles rather than to
original sources.)

Considering frames of observers related through Lorentz
transformations we have shown that entanglement is
frame-independent but the violation of Bell's inequality is
frame-dependent. ( Compare Refs. 4-8.)  We show that analogous features arise as the spin
undergoes precession in a magnetic field. Correlations are
formulated for the above cases. We note how, though, the spins
rotate in a magnetic field, time-dependence disappears for the
entangled state, consistently with its total spin zero.

The effects of a Lorentz transformation is uniform in space-time.
A magnetic field introduces time-dependence through spin rotations
but is homogenous in space.

What happens when particles created in an entangled state find
themselves plunged in a space-time dependent external field? Among
other possible consequences the spin  correlations can depend on
the location in space and time of the measurements carried out.
Does the entanglement survive uniformly?

The Dirac equation (generalized to include terms corresponding to
anomalous magnetic moment and even electric dipole moment) has
been solved for a class of plane wave fields which includes plane
-polarized ones. (See ref.9 and sources cited there.) Apart from
special features arising from 4-component Dirac spinors involved,
more fundamentally, space-time dependence is present.

Spinors have been studied in black hole metrics. A standard
reference is Sec. 10 of Ref.11. Kerr-Schild formalism is
elucidated in Ref.12. Entanglement beyond special relativity has
been considered in Ref.6. In such contexts how close can one bring
together the mysteries of entanglement and black holes? What
happens when entangled particles start moving away along
geodesics? We hope to understandsuch aspects better in future.

\begin{appendix}

\section{Lorentz transformations and Wigner rotations of spin}
\setcounter{equation}{0}

The full treatment of Wigner rotations of polarization, the
expectation values of the canonical spin operator, can be found in
Ref. 9. We present below briefly the essential results (based on
equations (2.24-34) of Ref. 9). Let $u$ be the 4-velocity of a
particle, of mass $m$ and spin $s$, in a particular frame. A pure
Lorentz transformation corresponding to a 4-velocity $u''$ will
give one, $u'$, with
\begin{equation}
u'_0=u_0u_0''+\overrightarrow{u}\cdot \overrightarrow{u}''.
\end{equation}
Define
\begin{equation}
a=\left(1+u_0\right)\left(1+u_0''\right)\left(1+u_0'\right),\qquad
b=1+u_0+u_0''+u'_0.
\end{equation}
The momentum and the polarization, both turn about the axis
$\overrightarrow{u}\times \overrightarrow{u}''$, but through
angles $\alpha$ and $\delta$ respectively where
\begin{equation}
\cos^2\frac \delta 2=\frac {b^2}{2a}
,\qquad\cos\left(\alpha-\delta\right)=\frac{u_0u_0'-u_0''}{\left|\overrightarrow{\textbf{u}}\right|\left|\overrightarrow{\textbf{u}}'\right|}
=\frac{p_0p_0'-mp_0''}{\left|\overrightarrow{\textbf{p}}\right|\left|\overrightarrow{\textbf{p}}'\right|}
\end{equation}
This displays the "lag" mentioned in the Introduction.  Evidently
this lag vanishes as $m\rightarrow 0$. One can also write
\begin{equation}
\sin\delta=\frac ba \left|\overrightarrow{\textbf{u}}\times
\overrightarrow{\textbf{u}}''\right|,\qquad\cos\delta=1-\frac 1a
\left|\overrightarrow{\textbf{u}}\times
\overrightarrow{\textbf{u}}''\right|^2,
\end{equation}
This displays explicitly the fact that for
\begin{equation}
\overrightarrow{\textbf{u}}\times
\overrightarrow{\textbf{u}}''=0,\qquad \delta=0.
\end{equation}
Thus, in our canonical formulation, the polarization does not
rotate under a pure Lorentz transformation parallel (or
antiparallel) to the initial momentum. [Hence, the prescription of
defining the polarization by passing to the rest frame is totally
superfluous in our canonical formalism. One defines it directly in
any frame, introducing canonical spin matrices.].

For a spin $\frac 12$ particle a rotation $\delta$ about an axis
$\widehat{\textbf{k}}$ (say, with $\widehat{\textbf{k}}^2=1$)
transforms the basis states $\left|\pm\right\rangle$ as, with
$\overrightarrow{\sigma}$ denoting Pauli matrices
\begin{equation}
e^{\mathbf{i}\frac \delta
2\widehat{\textbf{k}}\cdot\overrightarrow{\sigma}}\left(\begin{matrix}
  \left|+\right\rangle \\
  \left|+\right\rangle  \\
\end{matrix}\right)=\left(\begin{matrix}
  \left(\cos\frac\delta 2+\mathbf{i}\widehat{k}_3\sin\frac\delta 2\right)\left|+\right\rangle+
  \mathbf{i}\left(\widehat{k}_1-i\widehat{k}_2\right)\sin\frac\delta 2 \left|-\right\rangle\\
  \mathbf{i}\left(\widehat{k}_1+\mathbf{i}\widehat{k}_2\right)\sin\frac\delta 2
  \left|+\right\rangle+ \left(\cos\frac\delta 2-\mathbf{i}\widehat{k}_3\sin\frac\delta 2\right)\left|-\right\rangle \\
\end{matrix}\right).
\end{equation}

\section{Precession of polarization in a constant magnetic field}
\setcounter{equation}{0}

The Thomas equation for precession of polarization was derived in
our previous papers (cited in Ref. 9) starting with Wigner
rotations of canonical spin. Solutions were presented. Here we
present the solutions in a fashion compact and well-adapted to our
present goal. The key is the choice of axes displaying clearly the
role of initial conditions, with study of correlations in view.
Let $\overrightarrow{\textbf{B}}$ be the constant magnetic field
and $\overrightarrow{\textbf{v}}$ the velocity of the positive
$m$, spinning particle. Let
$\overrightarrow{\textbf{B}}=B\widehat{\textbf{B}}$,
$\overrightarrow{\textbf{v}}=v\widehat{\textbf{v}}$
($\widehat{\textbf{B}}^2=\widehat{\textbf{v}}^2=1$) (We use
standard units with $c=1$, $v<1$.) Two constants of motion are $v$
and $\widehat{\textbf{B}}\cdot \widehat{\textbf{v}}$. we start
with the general case
\begin{equation}
0\leq \left(\widehat{\textbf{B}}\cdot
\widehat{\textbf{v}}\right)^2<1.
\end{equation}
The limiting case
\begin{equation}
\widehat{\textbf{B}}\times \widehat{\textbf{v}}=0, \qquad
\widehat{\textbf{B}}\cdot \widehat{\textbf{v}}=\pm 1.
\end{equation}
is particularly simple and best treated separately. The case
\begin{equation}
\widehat{\textbf{B}}\cdot \widehat{\textbf{v}}=0
\end{equation}
is also simple, but contained in (B.1). We choose, for (B.1), the
ortho-normalized fixed axes
\begin{equation}
\widehat{\textbf{B}}, \qquad
\frac{\left(\widehat{\textbf{B}}\times
\widehat{\textbf{v}}\right)_{(0)}}{\sqrt{1-\left(\widehat{\textbf{B}}\cdot
\widehat{\textbf{v}}\right)^2}},\qquad
\frac{\widehat{\textbf{B}}\times\left(\widehat{\textbf{B}}\times
\widehat{\textbf{v}}\right)_{(0)}}{\sqrt{1-\left(\widehat{\textbf{B}}\cdot
\widehat{\textbf{v}}\right)^2}},
\end{equation}
where $(0)$ denotes the initial value at $t=0$. The corresponding
components of the polarization $\overrightarrow{\Sigma}$ are
defined as
\begin{equation}
\Sigma_1=\widehat{\textbf{B}}\cdot \overrightarrow{\Sigma},\qquad
\Sigma_2=\frac{\left(\widehat{\textbf{B}}\times
\widehat{\textbf{v}}\right)_{(0)}\cdot\overrightarrow{\Sigma}
}{\sqrt{1-\left(\widehat{\textbf{B}}\cdot
\widehat{\textbf{v}}\right)^2}},\qquad
\Sigma_3=\frac{\left(\widehat{\textbf{B}}\times\left(\widehat{\textbf{B}}\times
\widehat{\textbf{v}}\right)_{(0)}\right)\cdot
\overrightarrow{\Sigma}}{\sqrt{1-\left(\widehat{\textbf{B}}\cdot
\widehat{\textbf{v}}\right)^2}},
\end{equation}
Next one defines the ortho-normalized rotating set of axes (with
values at time $t$)
\begin{equation}
\widehat{\textbf{B}}, \qquad
\frac{\left(\widehat{\textbf{B}}\times
\widehat{\textbf{v}}\right)}{\sqrt{1-\left(\widehat{\textbf{B}}\cdot
\widehat{\textbf{v}}\right)^2}},\qquad
\frac{\widehat{\textbf{B}}\times\left(\widehat{\textbf{B}}\times
\widehat{\textbf{v}}\right)}{\sqrt{1-\left(\widehat{\textbf{B}}\cdot
\widehat{\textbf{v}}\right)^2}},
\end{equation}
The rotation is also about $\widehat{\textbf{B}}$, given by
$\frac{d\widehat{\textbf{v}}}{dt}=-\overrightarrow{\omega}\times
\widehat{\textbf{v}}$ with
\begin{equation}
\overrightarrow{\omega}=\frac{eB}{m\gamma}\widehat{\textbf{B}}
\qquad \gamma=\left(1-v^2\right)^{-1/2}.
\end{equation}
One has (with $\overrightarrow{\omega}=\omega\widehat{\omega}$,
$\widehat{\omega}^2=1$)
\begin{eqnarray}
&&\left(\widehat{\textbf{B}}\times\widehat{\textbf{v}}\right)=\left(\widehat{\textbf{B}}\times\widehat{\textbf{v}}\right)_{(0)}\cos\omega
t-\left(\widehat{\textbf{B}}\times
\left(\widehat{\textbf{B}}\times
\widehat{\textbf{v}}\right)\right)_{(0)}\sin\omega t,\nonumber\\
&&\left(\widehat{\textbf{B}}\times
\left(\widehat{\textbf{B}}\times
\widehat{\textbf{v}}\right)\right)=\left(\widehat{\textbf{B}}\times
\left(\widehat{\textbf{B}}\times
\widehat{\textbf{v}}\right)\right)_{(0)}\cos\omega
t+\left(\widehat{\textbf{B}}\times\widehat{\textbf{v}}\right)_{(0)}\sin\omega
t
\end{eqnarray}
One defines correspondingly the rotating components of
$\overrightarrow{\Sigma}^{(r)}$ as
\begin{equation}
\Sigma_1^{(r)}=\Sigma_{1},\qquad \Sigma_2^{(r)}=\Sigma_{2}\cos
\omega t-\Sigma_{3}\sin \omega t,\qquad
\Sigma_3^{(r)}=\Sigma_{3}\cos \omega t+\Sigma_{2}\sin \omega
t.\end{equation} The precession equation for
$\overrightarrow{\Sigma}$ is
\begin{equation}
\frac{d\overrightarrow{\Sigma}}{dt}=-\left(\overrightarrow{\omega}+\overrightarrow{\Omega}
\right)\times \overrightarrow{\Sigma},
\end{equation}
where
\begin{equation}
\overrightarrow{\Omega}=\frac{\alpha
eB}{m\gamma}\left(\gamma\widehat{\textbf{B}}-\left(\gamma-1\right)\left(\widehat{\textbf{B}}\cdot
\widehat{\textbf{v}}\right)\widehat{\textbf{v}}\right)
\end{equation}
with $\alpha=(g-2)/2$ representing the effect of the anomalous
magnetic moment. Carefully grouping the terms, after
simplifications, one indeed obtains the expected result
\begin{equation}
\frac{d\overrightarrow{\Sigma}^{(r)}}{dt}=-\overrightarrow{\Omega}
\times \overrightarrow{\Sigma}^{(r)}.
\end{equation}
The effect of $\overrightarrow{\omega}$ is absorbed in the
rotating frame. One has just the supplementary rotation $\Omega t$
about $-\widehat{\Omega}$ (where
$\overrightarrow{\Omega}=\Omega\;\widehat{\Omega}$,
$\widehat{\Omega}^2=1$). In fact, since
\begin{equation}
\left(\widehat{\textbf{B}}\times\widehat{\textbf{v}}\right)\cdot\widehat{\Omega}=0
\end{equation}
one obtains for $\overrightarrow{\Sigma}^{(r)}$
\begin{equation}
\frac{d{\Sigma_1^{r)}}}{dt}=\Omega_3\Sigma_2^{(r)},\qquad
\frac{d{\Sigma_3^{r)}}}{dt}=-\Omega_1\Sigma_2^{(r)},\qquad
\frac{d{\Sigma_2^{r)}}}{dt}=\Omega_1\Sigma_3^{(r)}-\Omega_3\Sigma_1^{(r)}
\end{equation}
with the constant coefficients
\begin{eqnarray}
&&\Omega_1=\widehat{\mathbf{B}}\cdot
\overrightarrow{\Omega}=\frac{\alpha
eB}{m\gamma}\left(\gamma-\left(\gamma-1\right)\left(\widehat{\textbf{B}}\cdot
\widehat{\textbf{v}}\right)^2\right),\nonumber\\
&&\Omega_2=\frac{\left(\widehat{\textbf{B}}\times
\widehat{\textbf{v}}\right)\cdot
\overrightarrow{\Omega}}{\sqrt{1-\left(\widehat{\textbf{B}}\cdot
\widehat{\textbf{v}}\right)^2}}=0,\nonumber\\
&&\Omega_3=\frac{\left(\overrightarrow{\mathbf{B}}\times\left(\widehat{\textbf{B}}\times
\widehat{\textbf{v}}\right)\right)\cdot
\overrightarrow{\Omega}}{\sqrt{1-\left(\widehat{\textbf{B}}\cdot
\widehat{\textbf{v}}\right)^2}}=\frac{\alpha
eB}{m\gamma}\left(\gamma-1\right)\left(\widehat{\textbf{B}}\cdot
\widehat{\textbf{v}}\right)\sqrt{1-\left(\widehat{\textbf{B}}\cdot
\widehat{\textbf{v}}\right)^2}
\end{eqnarray}
with
\begin{equation}
\Omega_1^2+\Omega_2^2+\Omega_3^2=\left(\frac{\alpha
eB}{m}\right)^2\left(1-\left(\widehat{\textbf{B}}\cdot
\widehat{\textbf{v}}\right)^2\right)=\Omega^2.
\end{equation}
The explicit expressions for $\left(\Omega_1,\Omega_3\right)$ are
not particulary simple. But they are precisely what are needed to
display the basic structure of the solutions.

The linear set (B.14) with constant coefficient is easily solved.
One obtains
\begin{eqnarray}
&&\Sigma_1^{(r)}=\frac{\Omega_3}{\Omega}\left(a\cos\Omega
t+b\sin\Omega
t\right)+c\left(\frac{\Omega_1}{\Omega}\right),\nonumber\\
&&\Sigma_3^{(r)}=-\frac{\Omega_1}{\Omega}\left(a\cos\Omega
t+b\sin\Omega
t\right)+c\left(\frac{\Omega_3}{\Omega}\right),\nonumber\\
&&\Sigma_2^{(r)}=\left(b\cos\Omega t-a\sin\Omega t\right),
\end{eqnarray}
where (since $\overrightarrow{\Sigma}_{(0)}^{(r)}
=\overrightarrow{\Sigma}_{(0)}$)
\begin{equation}
a=\frac{1}{\Omega}\left(\Omega_3\Sigma_{1(0)}-\Omega_1\Sigma_{3(0)}\right),\qquad
c=\frac{1}{\Omega}\left(\Omega_1\Sigma_{1(0)}+\Omega_3\Sigma_{3(0)}\right),\qquad
b=\Sigma_{2(0)}
\end{equation}
and hence
\begin{equation}
\left|\overrightarrow{\Sigma}^{(r)}\right|^2=a^2+b^2+c^2=\left|\overrightarrow{\Sigma}^{(r)}_{(0)}\right|^2.
\end{equation}
Finally one obtains the expected, standard, form
\begin{equation}
\overrightarrow{\Sigma}^{(r)}=\left(\cos\Omega t\right)
\overrightarrow{\Sigma}_{(0)}+\left(1-\cos\Omega t\right)
\left(\widehat{\Omega}\cdot\overrightarrow{\Sigma}\right)_{(0)}\widehat{\Omega}-\left(\sin\Omega
t\right)
\left(\widehat{\Omega}\times\overrightarrow{\Sigma}\right)_{(0)}.
\end{equation}
This represents a rotation $\left(\Omega t\right)$ about the axis
$\left(-\widehat{\Omega}\right)$. Trivially inverting (B.8) one obtains the projections on fixed axes. Thus we have two
successive rotations of
\begin{enumerate}
    \item angle $\left(\omega t\right)$ about the axis
$\left(-\widehat{\omega}\right)$
    \item angle $\left(\Omega t\right)$ about the axis
$\left(-\widehat{\Omega}\right)$
\end{enumerate}

\paragraph {\bf The limit $\left(\widehat{\mathbf{B}}\cdot
\widehat{\mathbf{v}}\right)=0$:} In this case the two axes
coincide. One has just a rotation $\left(\Omega+\omega\right)t$
about $\left(-\widehat{\mathbf{B}}\right)$ where
\begin{equation}
\overrightarrow{\Omega}=\alpha \gamma\overrightarrow{\omega}.
\end{equation}
Now the axes (B.5) reduce to
\begin{equation}
\left(\widehat{\mathbf{B}},\widehat{\mathbf{B}}\times
\widehat{\mathbf{v}},-\widehat{\mathbf{v}}\right)
\end{equation}
and one has
\begin{eqnarray}
&&\left(\widehat{\textbf{B}}\times\widehat{\textbf{v}}\right)=\left(\widehat{\textbf{B}}\times\widehat{\textbf{v}}\right)_{(0)}\cos\omega
t+\widehat{\textbf{v}}_{(0)}\sin\omega t,\nonumber\\
&&\widehat{\textbf{v}}=\widehat{\textbf{v}}_{(0)}\cos\omega
t-\left(\widehat{\textbf{B}}\times\widehat{\textbf{v}}\right)_{(0)}\sin\omega
t
\end{eqnarray}

\paragraph {\bf The special limit
$\left(\widehat{\mathbf{B}}\times \widehat{\mathbf{v}}\right)=0$:}
Now
\begin{equation}
1-\left(\widehat{\mathbf{B}}\cdot \widehat{\mathbf{v}}\right)^2=0
\end{equation}
and one cannot normalize as in (B.5) and (B.6). But now
$\widehat{v}$ remains constant and
\begin{equation}
\overrightarrow{\omega}=\frac{eB}{m\gamma}\widehat{\mathbf{B}},\qquad
\overrightarrow{\Omega}=\alpha \overrightarrow{\omega}.
\end{equation}
(Note the difference of a factor $\gamma$ for
$\overrightarrow{\Omega}$, as compared to (B.21).) Choosing any
pair of fixed ortho-normal axes in the plane orthogonal to
$\overrightarrow{\mathbf{B}}$ one obtains quite simply the final
results.
\end{appendix}

\end{document}